# Anomalous Collision of Exceptional Points on Nonorientable Manifolds

Weijia Wang,[†] Qicheng Zhang,[†*] Kun Zhang, Shuaishuai Tong, and Chunyin Qiu[*]

*Key Laboratory of Artificial Micro- and Nano-Structures of Ministry of Education and School of Physics and Technology, Wuhan University, Wuhan 430072, China*

[†]*These authors contributed equally: Weijia Wang, Qicheng Zhang*
[*]*To whom correspondence should be addressed: qicheng.zhang@whu.edu.cn; cyqiu@whu.edu.cn*

Band degeneracies, ranging from Hermitian Dirac points to non-Hermitian exceptional points (EPs), play a central role in topological phase transitions. Beyond the topology of individual degeneracies, their mutual interactions yield richer phenomena. A representative example is the anomalous non-annihilating collision of pairwise-created degeneracies, previously believed to occur only in non-Abelian multiband systems. Here, we theoretically reveal and experimentally demonstrate that such an anomalous collision can emerge even in a simple two-band system without non-Abelian nature. In a two-dimensional non-Hermitian lattice whose Brillouin zone forms a nonorientable Klein bottle, two EPs with opposite topological charges—pairwise created from a hybrid point—merge into a new vortex point upon re-encounter, instead of annihilating. Remarkably, the hybrid point is a defective degeneracy featuring no eigenenergy braiding, whereas the vortex point is a non-defective degeneracy yet exhibits nontrivial eigenenergy braiding. This process manifests a non-Hermitian phase transition from a gapped phase to a gapless phase, a scenario that we directly observe in a hybrid-dimensional acoustic lattice via momentum-resolved band braid and Berry phase measurements. Our findings identify nonorientability as a new arena for engineering band degeneracies and topological phases, and pave the way for experimentally exploring the interplay between exceptional and nonorientable topology.

Band degeneracies play a central role in modern condensed matter physics. In Hermitian systems, prototypical band degeneracies such as Dirac points and Weyl points provide the backbone for a broad family of topological phases, including topological insulators and topological semimetals [1-4]. These Hermitian degeneracies, involving the coalescence of eigenenergies only, are typically protected by symmetries and characterized by quantized Berry phases. In contrast, non-Hermitian systems give rise to qualitatively distinct classes of degeneracies, known as exceptional points (EPs), at which both eigenenergies and eigenstates coalesce [5-10]. The EPs underpin a variety of phenomena without Hermitian counterparts, including complex band braiding [11-23], chiral state transfer [24-28], enhanced sensing [29-32] and perfect absorption [33-36].

Beyond the topology involved in individual band degeneracies, interactions of multiple degeneracies enable a much richer landscape [37-45]. In two-band systems defined on a conventional Brillouin zone (BZ), the fermion doubling theorem [37,41] dictates that pairwise-created degeneracies with opposite topological charges must ultimately annihilate upon collision, as illustrated in Fig. 1(a). In contrast, in multiband systems, the no-go theorem [42] and non-Abelian conservation rule [43] allow for a more striking scenario: pairwise-created degeneracies need not annihilate when they collide again, because interacting with other degeneracies can transmute their topological charges, as sketched in Fig. 1(b). Such anomalous collisions have been extensively studied and experimentally verified, both for Hermitian Dirac points protected by quaternion invariants [46-49] and for non-Hermitian EPs characterized by non-Abelian braiding invariants [43,50,51]. Nevertheless, all these observations arise from the interplay between band degeneracies and the topology of an orientable space, typically a torus BZ or an infinite parameter space. Recent theoretical advances have revealed that nonsymmorphic symmetries can render momentum space nonorientable, leading to BZs with the topology of a Klein bottle or a real projective plane [52-60]. Such a fundamental change reshapes the topological classification, invalidates the conventional fermion-doubling theorem, and establishes new no-go constraints [58-60]. This naturally raises a key question: how does nonorientability affect the interaction of EPs, and can anomalous collisions occur without relying on multiband non-Abelian topology?

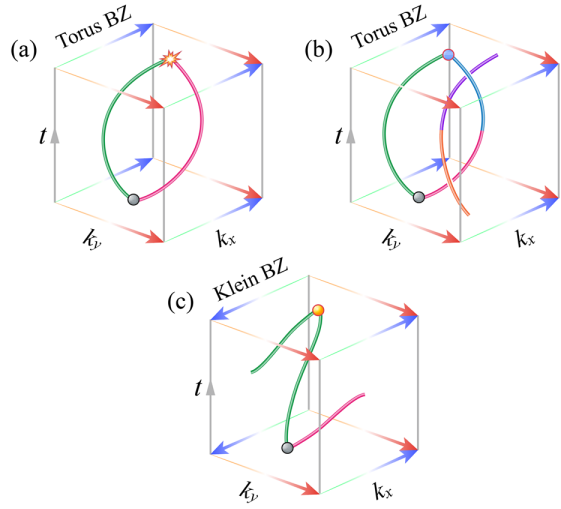

FIG. 1. Representative band-degeneracy evolutions in different scenarios. (a) In a two-band Abelian system, pairwise-created degeneracies annihilate upon collision on an orientable torus BZ. (b) In a multiband non-Abelian system, pairwise-created degeneracies can merge into a new degeneracy without annihilation. (c) In a two-band Abelian system, pairwise-created degeneracies can also merge into a new degeneracy without annihilation, but on a nonorientable Klein BZ. Opposite arrows along the $k_x$ direction signify the nonorientability of the Klein BZ.

Here, we answer this question by revealing a nonorientability-induced anomalous EP evolution in a two-dimensional (2D) two-band lattice whose fundamental domain of the BZ forms a Klein bottle. As depicted in Fig. 1(c), two EPs with opposite topological charges, initially pairwise created from a hybrid point (HP), merge into a distinct vortex point (VP) upon re-encounter, instead of annihilating. This anomalous behavior, unique to two-band systems defined on nonorientable manifolds, originates from the topological charge inversion of an EP when it traverses the nonorientable boundary of the Klein BZ (KBZ). Remarkably, although both the HP and VP are intrinsically tied to EP pairs, the former is a (non-Hermitian-type) defective degeneracy featuring no eigenenergy braiding, whereas the latter is a (Hermitian-type) non-defective degeneracy yet exhibits nontrivial eigenenergy braiding [12,61]. Consequently, this anomalous evolution signals a topological phase transition from a trivial gapped phase to a nontrivial gapless phase, the latter being identified by a monopole that is forbidden in both Hermitian systems and non-Hermitian systems defined on orientable BZs [60]. To experimentally evidence these effects, we implement a mixed-dimensional non-Hermitian acoustic lattice composed of one real-space dimension and one synthetic-momentum dimension. By measuring two independent yet complementary topological invariants—namely, the eigenfrequency braids and eigenstate Berry phases—along the KBZ boundary at four representative stages, we unambiguously observe the gapped-gapless phase transition associated with anomalous EP evolution, together with the existence of HP and VP. These experimental results further shed light on the validity of no-go theorem and the violation of fermion doubling theorem within nonorientable spaces.

**Theoretical model**

We construct a 2D two-band non-Hermitian lattice model to reveal anomalous EP collision on a nonorientable KBZ. As shown in Fig. 2(a), the model consists of two sublattices A and B per unit cell, and features reciprocal constant couplings $s_1$ and $s_2$, together with unidirectional time-dependent couplings $g_1(t)$ and $-g_1(t)$. The lattice Hamiltonian in 2D momentum space is expressed as

$$\mathbf{H}(k_x, k_y) = (s_2 \cos k_x + s_1)\sigma_x + (s_2 \sin k_x - ig_1(t)e^{ik_y})\sigma_y, \quad (1)$$

where $\sigma_x$ and $\sigma_y$ are Pauli matrices. Specifically, we set the parameters $s_1 = -0.6$, $s_2 = 1$, and $g_1(t) = te^{-i\pi/3}$, such that the Hamiltonian satisfies a momentum-space glide-reflection symmetry

$$U\mathbf{H}(k_x, k_y)U^{-1} = \mathbf{H}(-k_x, k_y + \pi), \quad (2)$$

with the unitary operator $U = \sigma_x$ (see Supplemental Information for details of the model design). Owing to this symmetry, it suffices to consider a reduced fundamental domain of BZ, $(k_x, k_y) \in [-\pi, \pi] \times [0, \pi]$, to fully capture the band topology of the lattice. Unlike the conventional BZ, $(k_x, k_y) \in [-\pi, \pi] \times [-\pi, \pi]$, that forms an orientable torus when its periodic boundaries are glued together, this reduced BZ corresponds to a nonorientable Klein bottle, referred to as the KBZ. Notably, the boundaries $k_y = 0$ and $k_y = \pi$ must be glued together in opposite directions due to the symmetry $U\mathbf{H}(k_x, 0)U^{-1} = \mathbf{H}(-k_x, \pi)$, thereby forming the nonorientable boundaries of the KBZ.

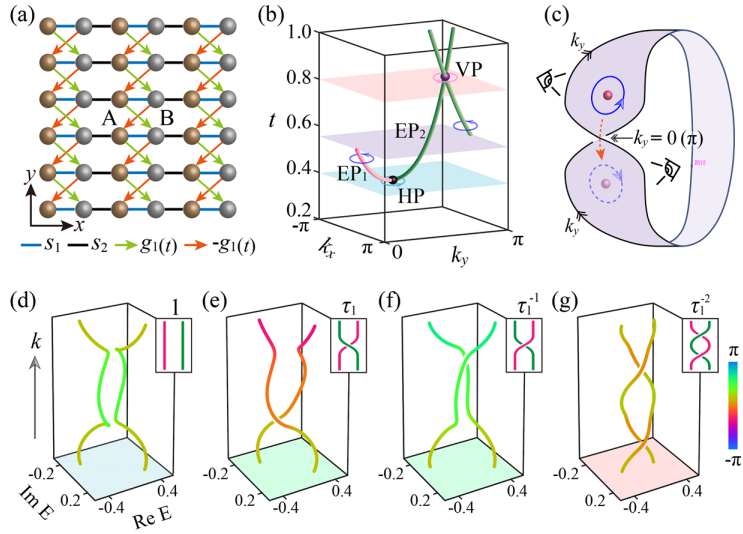

FIG. 2. Theoretical model for anomalous EP collision. (a) 2D non-Hermitian lattice model, with $s_1$ and $s_2$ representing reciprocal constant couplings, and $g_1(t)$ and $-g_1(t)$ representing time-dependent unidirectional couplings. (b) Adiabatic evolution of band degeneracies on the KBZ as a function of time $t$. The cyan, purple, and pink planes label the creation of a HP, the crossing of EP$_1$ through nonorientable boundaries, and the formation of a VP, respectively. (c) Möbius structure formed by two nonorientable boundaries $k_y = 0$ and $k_y = \pi$, illustrating the inversion of topological charge before and after EP$_1$ crosses the boundaries. (d)-(g) Eigenenergy braids (lines) and Berry phase accumulations (color) calculated along the counterclockwise loops in (b), encircling the HP, EP$_1$, EP$_2$, and VP, respectively. The insets show schematic braid diagrams.

We calculate the adiabatic evolution of band degeneracies on the KBZ as time $t$ varies, as shown in Fig. 2(b). To identify the exceptional classification of each degeneracy, we characterize them using both the topological charge (also referred to as the braiding invariant) defined via eigenenergy braids [14-16], and the Berry phase derived from biorthogonal eigenstates [62,63]. Specifically, the topological charge is assigned by the Artin word (e.g., $\tau_1$, $\tau_1^{-1}$ and 1) of the braid group, defined from the eigenenergy braid along a local counterclockwise loop $\Gamma_1$ encircling the degeneracy, along which the momenta return to themselves. Meanwhile, the loop is chosen to avoid crossing the nonorientable boundaries, so that all EP charges are defined with respect to loops lying on the same side of the Möbius strip formed by the twisted gluing of the $k_y = 0$ and $k_y = \pi$ boundaries, as illustrated in Fig. 2(c). In contrast, the Berry phase is defined as

$$\theta = i \oint_{\Gamma_2} \frac{\langle \phi(k)|\partial_k \psi(k)\rangle}{\langle \phi(k)|\psi(k)\rangle} dk, \tag{3}$$

where $\Gamma_2$ is a momentum loop along which the eigenstates return to themselves, and $\langle \phi(k)|$ and $|\psi(k)\rangle$ denote the left and right Bloch eigenstates, respectively. Notably, $\Gamma_1$ and $\Gamma_2$ need not coincide, as the momenta returning to themselves do not necessarily imply that the eigenstates also return to themselves. Remarkably, the eigenenergy braid and the Berry phase constitute independent yet complementary invariants, whose combination leads to distinct classifications of the band degeneracies considered in our work.

At moment $t = 0.4$, a band degeneracy emerges at $(k_x, k_y) = (0, \pi/3)$. As shown in Fig. 2(d), its trivial topological charge 1 and zero Berry phase identify it as a HP—a special type of non-Hermitian defective degeneracy distinct from conventional EPs, which typically carry nontrivial topological charge and nonzero Berry phase [63]. Upon increasing $t$, this HP splits into a pair of degeneracies. As shown in Fig. 2(e), the two degeneracies exhibit nontrivial topological charges, $\tau_1$ and $\tau_1^{-1}$, together with the Berry phase of $-\pi$, confirming that they are conventional EPs, referred to as EP$_1$ and EP$_2$, respectively. (Note that the color variation in Fig. 2(e) only reflects a phase accumulation of $-\pi/2$ after a single momentum loop $\Gamma_1$, during which the two eigenstates are exchanged; a double cyclic encircling, i.e., $\Gamma_2 = 2\Gamma_1$, is required for each eigenstate to return to itself, yielding the Berry phase $-\pi$.) At $t = 0.55$, the EP$_1$ carrying charge $\tau_1$ transverses the nonorientable boundary $k_y = 0$ and re-enters the KBZ from $k_y = \pi$, after which its charge is inverted to $\tau_1^{-1}$. This inversion can be intuitively understood from the Möbius trip formed by $k_y = 0$ and $k_y = \pi$, as shown in Fig. 2(c). Before crossing the nonorientable boundaries, EP$_1$ lies on the outer side of the Möbius strip with its encircling loop oriented counterclockwise (from the outer-side viewpoint). After crossing, EP$_1$ moves to the inner side, with its loop remaining counterclockwise (from the inner-side viewpoint). Since EP charges are defined with respect to loops lying on the same side of the Möbius strip, a counterclockwise loop on the inner side must be viewed as a clockwise loop on the outer side, which ultimately leads to the inversion of the EP topological charge. Finally, at $t = 0.8$, EP$_1$ and EP$_2$, both carrying identical topological charges $\tau_1^{-1}$, merge into a new degeneracy at $(k_x, k_y) = (-3\pi/10, 5\pi/6)$. This degeneracy hosts a nontrivial topological charge $\tau_1^{-2}$ but a zero Berry phase, as shown in Fig. 2(f), identifying it as a VP—a special Hermitian non-defective degeneracy distinct from conventional Dirac points that carry trivial topological charges in terms of eigenenergy braiding and nonzero Berry phase [1,2]. The VP also constitutes an unpaired monopole with total braiding degree two, which is forbidden in orientable non-Hermitian systems. This full evolution reveals that pairwise-created EPs do not annihilate upon re-encounter, but instead merge into a new degeneracy protected by the nonorientable topology, driving a transition from a gapped (trivial band braid 1) phase to a gapless phase (nontrivial band braid $\tau_1^{-2}$). For comparison, we also construct a trivial evolution scenario in which the EP crosses the orientable boundary of the KBZ. In that case, two EPs created from a HP recombine into the original one upon re-encounter, and the system remains in a trivial gapped phase throughout (see Supplemental Information).

**Experimental implementations**

We experimentally implement a mixed-dimensional acoustic lattice—comprising one real-space dimension and one synthetic-momentum dimension—to demonstrate the anomalous evolution of EPs. This implementation leverages the fact that our 2D momentum-space Hamiltonian $\mathbf{H}(k_x, k_y)$ can be mapped onto a 1D real-space Hamiltonian parameterized by the synthetic momentum $k_y$, i.e.,

$$\mathbf{H}(k_y) = \sum_n \left(g_1 e^{ik_y} + s_1\right)c_{2n}^\dagger c_{2n-1} + \left(-g_1 e^{ik_y} + s_1\right)c_{2n-1}^\dagger c_{2n} + s_2\left(c_{2n+1}^\dagger c_{2n} + c_{2n}^\dagger c_{2n+1}\right), (4)$$

where $c_n^\dagger$ and $c_n$ denote the creation and annihilation operators at the $n$th site, respectively. As shown in Fig. 3(a), the acoustic lattice consists of 16 acoustic cavities (8 unit cells) arranged along the real-space $x$ direction, forming a ring to realize periodic boundary conditions. Each cavity, with a dipole resonance frequency $\omega_0 \approx 5212$ Hz and an intrinsic loss $\gamma_0 \approx 40$ Hz, emulates a lattice site. Narrow tubes between neighboring cavities generate reciprocal constant acoustic couplings, e.g., parallel tubes for negative couplings $s_1 \approx -30$ Hz and crossed tubes for positive couplings $s_2 \approx 50$ Hz. Meanwhile, electroacoustic unidirectional couplers, each composed of a microphone, an amplifier, a phase shifter, and a loudspeaker, are introduced between cavities [19]. They enable $k_y$-dependent unidirectional complex couplings $\kappa_r(k_y) = 50 g_1(t) e^{ik_y}$ and $\kappa_l(k_y) = -50 g_1(t) e^{ik_y}$, where the $t$-dependent amplitudes are controlled by the amplifiers and the synthetic momentum $k_y$ is encoded via the phase shifter (see Supplemental Information for further experimental details).

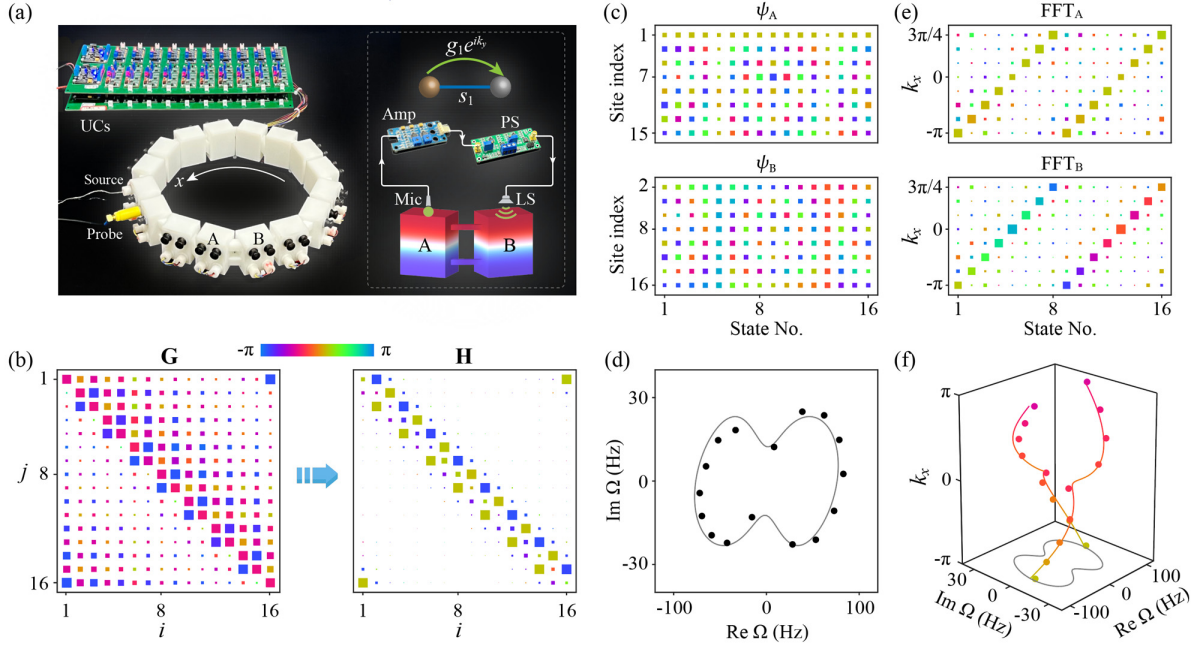

FIG. 3. Experimental implementation. (a) Acoustic lattice comprising one real-space dimension $x$ and one synthetic-momentum dimension $k_y$, where $k_y$ is encoded via unidirectional couplers (UCs), each consisting of a microphone (Mic), an amplifier (Amp), a phase shifter (PS), and a loudspeaker (Ls). (b) Left: Spatial representation of the experimentally measured Green's function at excitation frequency $\omega = 5212$ Hz. Bottom: Hamiltonian reconstructed from the measured Green's function. The square size and color indicate amplitude and phase, respectively. For clarity, the diagonal elements have been subtracted by the acoustic on-site energy $\omega_0 - i\gamma_0$. (c) Real-space eigenstates on sublattices A and B. (d) Eigenfrequencies of the system, with circles and solid lines representing experimental and theoretical results, respectively. (e) FFT spectra of corresponding eigenstates, where the position of maximum amplitude indicates the associated momentum $k_x$. (f) Eigenfrequency braid and Berry phase accumulation as a function of $k_x$. All experimental data in (b)-(f) are taken at $t = 0.48$ and $k_y = 0.25\pi$.

For a given moment $t$ and a synthetic momentum $k_y$, we sequentially excite and probe the 16 cavities to obtain the acoustic Green's function elements $\mathbf{G}_{ij}(\omega)$. Here, $i,j = 1:16$ label the probed and excited cavities, respectively, and $\omega$ denotes the excitation frequency. The left panel of Fig. 3(b) displays a representative spatial map of $\mathbf{G}(\omega)$ at $\omega = 5212$ Hz for given $t = 0.48$ and $k_y = 0.25\pi$ (corresponding to $\kappa_r = 23.18 - 6.21i$ Hz and $\kappa_l = -23.18 + 6.21i$ Hz). From the measured Green's function, the system Hamiltonian is experimentally reconstructed via $\mathbf{H} = \omega\mathbf{I} - \mathbf{G}^{-1}(\omega)$, where $\mathbf{I}$ is the identity matrix. In principle, $\mathbf{H}$ is independent of $\omega$, so the Green's function at any excitation frequency yields the same Hamiltonian. To suppress experimental errors, we average the reconstructed Hamiltonians over 51 frequencies uniformly sampled within $\omega \in [5162, 5262]$ Hz. As shown in the right panel of Fig. 3(b), the periodically distributed reciprocal and nonreciprocal nearest-neighbor couplings are clearly resolved, demonstrating the reliability of our experimental scheme. (Note that the diagonal elements of $\mathbf{H}$ have been subtracted by the acoustic onsite term $\omega_0 - i\gamma_0$ to highlight the coupling pattern.) From either the measured Green's function or the reconstructed Hamiltonian, we can obtain the real-space eigenstates $|\psi_n(x)\rangle$ and eigenfrequencies $\Omega_n$ of the system, with $n$ denoting the state index, as shown in Figs. 3(c) and 3(d). Notably, these quantities do not retain the momentum $k_x$ information. To retrieve the momentum, we apply a fast Fourier transform (FFT) to $|\psi_n(x)\rangle$. The real momentum $k_x$ of each eigenstate is then identified by the peak position of the FFT spectrum, as illustrated in Fig. 3(e). Accordingly, we extract the FFT amplitudes $\rho_{A,B}(k_x)$ and phases $\varphi_{A,B}(k_x)$ for sublattices A and B at each identified momentum $k_x$, reconstructing the momentum-resolved Bloch eigenstate $|\psi_n(k_x)\rangle = [\rho_A e^{i\varphi_A}, \rho_B e^{i\varphi_B}]^T$. The right eigenstate $|\psi_n(k_x)\rangle$ and its biorthogonal left counterpart $\langle\phi_n(k_x)|$ are then used to characterize the Berry phase, while the corresponding eigenfrequency $\Omega_n(k_x)$ constitutes the eigenvalue braids and topological charges. Figure 3(f) shows the eigenfrequency and Berry phase accumulation as $k_x$ varies from $-\pi$ to $\pi$ with a resolution of $0.25\pi$ (the Berry phase accumulation here has been slightly adjusted along the chosen path for data presentation). By further stepping the synthetic momentum $k_y$ from 0 to $\pi$ in increments of $0.25\pi$, we can obtain the full momentum-resolved eigenfrequencies and eigenstates across the entire KBZ.

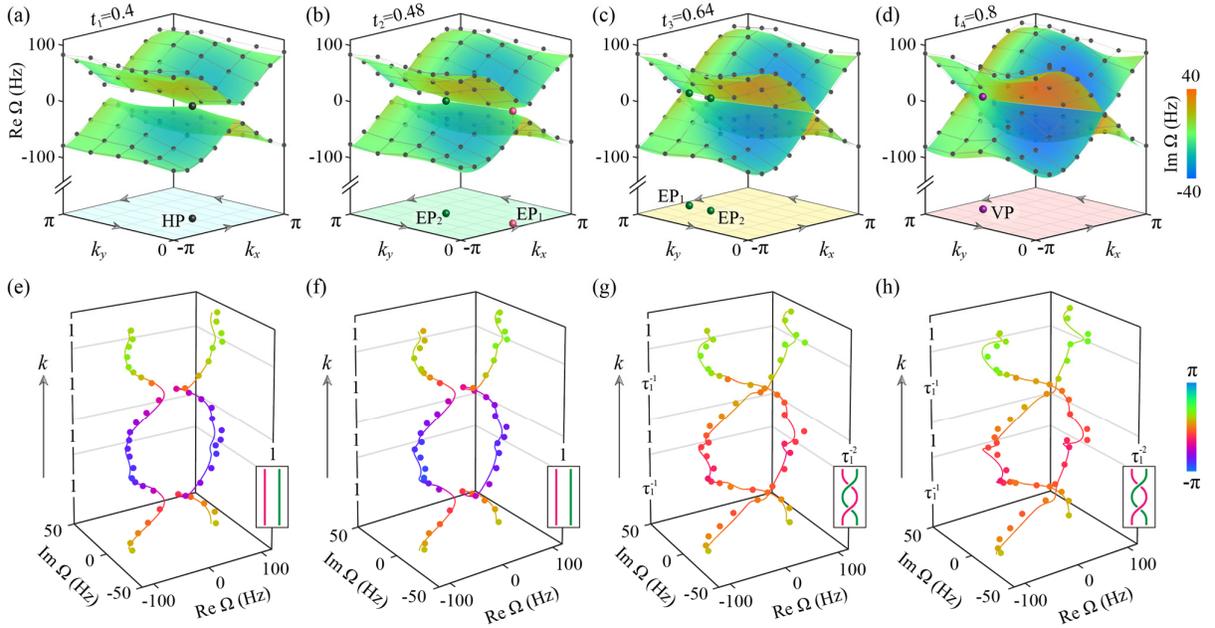

FIG. 4. Experimental characterization of the phase transition induced by anomalous EP evolution. (a)-(d) Measured eigenfrequencies (dots) over the entire KBZ at four representative moments $t_1 = 0.4$, $t_2 = 0.48$, $t_3 = 0.64$, and $t_4 = 0.8$, together with the corresponding theoretical Riemann surfaces. Spheres denote the band degeneracies. (e)-(h) Measured band braids (dots) and Berry phases (color) extracted from the experimental data along the KBZ boundary, with arrows indicating the traversal direction. Solid lines represent the theoretical results. The trivial band braids at $t_1$ and $t_2$, in contrast to the nontrivial band braids at $t_3$, and $t_4$, evidence a gapped-to-gapless topological phase transition.

First, we characterize the phase transition of the acoustic lattice induced by anomalous EP evolution. We focus on four representative moments $t_1 = 0.4$, $t_2 = 0.48$, $t_3 = 0.64$ and $t_4 = 0.8$, which correspond to the presence of a HP, a pair of EPs with opposite topological charges, a pair of EPs with identical topological charges, and a VP within the KBZ, respectively. These four moments are implemented by tuning the amplifiers in the electroacoustic feedback modules, such that the amplitudes of the unidirectional couplings in the acoustic lattice are fixed at $|\kappa_r| = |\kappa_l| \approx 20$ Hz, $|\kappa_r| = |\kappa_l| \approx 24$ Hz, $|\kappa_r| = |\kappa_l| \approx 32$ Hz, and $|\kappa_r| = |\kappa_l| \approx 40$ Hz. Using the strategy described above to access both the real momentum $k_x$ and the synthetic momentum $k_y$, the measured eigenfrequencies and the corresponding theoretical Riemann surfaces are presented in Figs. 4(a)-4(d), showing good agreement. We further extract the eigenfrequencies and eigenstates along the KBZ boundary to obtain the band braid and Berry phase at each moment, as shown in Figs. 4(e)-4(h). At $t_1$ and $t_2$, the band braids are trivial, i.e., 1, serving as a direct manifestation of a gapped phase [Figs. 4(e) and 4(f)]. Together with the zero Berry phase at both moments, these observations indicate the presence of either a single HP or a pair of EPs with opposite topological charges originating from the HP. In contrast, at $t_3$ and $t_4$, the band braids along the KBZ boundary become nontrivial, i.e., $\tau_1^{-2}$, signaling a gapless phase, while the Berry phase remains zero [Figs. 4(g) and 4(h)]. The two signatures imply the presence of a pair of EPs carrying identical topological charges, or equivalently, a VP formed by their merging. These results provide unambiguous experimental evidence for a topological phase transition driven by the anomalous evolution of EPs. They further verify the no-go theorem on the KBZ and demonstrate the violation of fermion-doubling theorem. By partitioning the KBZ boundary into four segments, we find that the braids at $t_3$ and $t_4$ can be expressed as $1\tau_1^{-1}1\tau_1^{-1}$, which satisfies the no-go constraint of the form $B_a B_b B_a B_b^{-1}$ [58,60]. (Note that the trivial braids at $t_1$ and $t_2$ also satisfies this structure.) The nonzero total discriminant number, e.g., braiding degree [44]—defined as the difference between the numbers of $\tau_1$ and $\tau_1^{-1}$—at $t_3$ and $t_4$ implies the existence of EPs that cannot annihilate and instead form unpaired monopole in a nonorientable momentum space, thereby violating the fermion-doubling theorem [58-60].

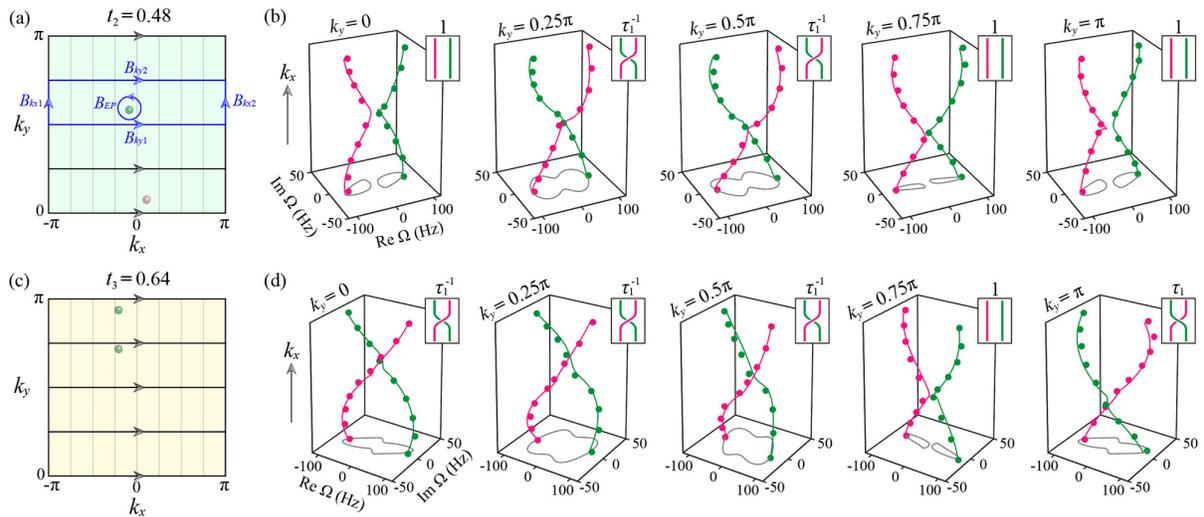

FIG. 5. Experimental characterization of EP charge inversion. (a) Schematic distribution of EPs on the KBZ at moment $t_2$. The local loop defining EP topological charge can be continuously deformed to four paths with constant $k_x$ and $k_y$. (b) Measured eigenfrequency braids along five constant-$k_y$ paths. The inequivalent braids 1 and $\tau_1^{-1}$ associated with the $k_y = 0.75\pi$ and $k_y = 0.5\pi$ paths indicate the presence of an EP carrying topological charge $\tau_1^{-1}$ between them, while the inequivalent braids $\tau_1^{-1}$ and 1 for the $k_y = 0.25\pi$ and $k_y = 0$ paths imply an EP with charge $\tau_1$. (c)-(d) Same as (a)-(b) but for moment $t_3$. The inequivalent braids $\tau_1$ and 1 for the $k_y = \pi$ and $k_y = 0.75\pi$ paths, together with 1 and $\tau_1^{-1}$ for the $k_y = 0.75\pi$ and $k_y = 0.5\pi$ paths, demonstrate that two EPs carrying topological charge $\tau_1^{-1}$ reside on the KBZ.

The transition from the gapped phase at $t_2$ to the gapless phase at $t_3$ necessarily implies that an EP transverses the nonorientable boundaries and undergoes a topological charge inversion. We next confirm this fact directly from our experimental results. To this end, we first recall an EP-detection theorem on the KBZ for two-band systems: if the eigenenergy braids associated with two neighboring constant-$k_y$ paths are inequivalent, i.e., $B_{k_{y1}} \neq B_{k_{y2}}$, then a nontrivial EP (or multiple EPs) carrying a (or total) topological charge $B_{k_{y2}}^{-1} B_{k_{y1}}$ must lie between these two paths [58]. This theorem can be proven from the loop homotopy of the EP charge $B_{EP} = B_{k_{x1}}^{-1} B_{k_{y2}}^{-1} B_{k_{x2}} B_{k_{y1}}$, which reduces to $B_{EP} = B_{k_{y2}}^{-1} B_{k_{y1}}$ by invoking the Abelian relation $B_{k_{y2}}^{-1} B_{k_{x2}} = B_{k_{x2}} B_{k_{y2}}^{-1}$ in two-band systems and the equivalence relation $B_{k_{x1}} = B_{k_{x2}}$ enforced by orientable boundaries, as illustrated in Fig. 5(a). Experimentally, at moment $t_2$, we measure the eigenfrequency braids along five constant-$k_y$ paths, i.e., $k_y = 0$, $0.25\pi$, $0.5\pi$, $0.75\pi$ and $\pi$, as shown in Fig. 5(b). The inequivalent braids $B_{k_y=0.25\pi} = \tau_1^{-1}$ and $B_{k_y=0} = \mathbb{I}$ indicate the presence of an EP with topological charge $\tau_1$ between the two paths. Similarly, the inequivalent braids $B_{k_y=0.75\pi} = 1$ and $B_{k_y=0.5\pi} = \tau_1^{-1}$ imply another EP carrying charge $\tau_1^{-1}$ in between. By contrast, at moment $t_3$, the inequivalent braids between the paths $k_y = \pi$ and $k_y = 0.75\pi$, namely $\tau_1$ and $\mathbb{I}$, together with those between $k_y = 0.75\pi$ and $k_y = 0.5\pi$, namely 1 and $\tau_1^{-1}$ [Figs. 5(c) and 5(d)], demonstrate that two EPs with the same topological charge $\tau_1^{-1}$ exist on the KBZ. These observations evidence that one EP crosses the nonorientable boundaries between $t_2$ and $t_3$, accompanied by a charge inversion. Additionally, our experimental results confirm a rule for boundary braid transformation: the braid along a boundary before and after an EP crosses it differ precisely by the EP's charge, i.e., $B_{after} = B_{before}B_{EP}^{-1}$ [60]. For instance, the braid along the boundary $k_y = 0$ in our system changes from 1 to $\tau_1^{-1}$, satisfying $B_{k_y=0}^{t_3} = B_{k_y=0}^{t_2}B_{EP}^{-1}$.

**Conclusion**

We have theoretically revealed and experimentally demonstrated an anomalous EP evolution on a nonorientable KBZ, where pairwise-created EPs from a HP do not annihilate upon collision but instead merge into a distinct VP. Our results highlight nonorientability as a fundamentally new ingredient for engineering band degeneracies and topological phases in non-Hermitian systems. Unlike previously studied anomalous collisions of degeneracies that rely on the non-Abelian topology of orientable spaces, the anomalous EP evolution reported here originates from a global constraint enforced by the nonorientable geometry of the KBZ. The inversion of EP topological charge upon traversing the nonorientable boundary provides a direct mechanism for inducing non-Hermitian gapped-gapless topological phase transition and for violating the fermion doubling theorem. In particular, the eigenfrequency braids and Berry phases measured in our experiments, as two independent yet complementary invariants, offer a unified probe for classifying exceptional band degeneracies like HPs and VPs. Such theoretical framework and experimental scheme can be naturally extended to explore richer degeneracy structures in multiband systems, where non-Abelian and higher-order EPs could undergo similarly constrained yet unconventional evolutions enforced by nonorientability [58,60].

*Note added.* We recently became aware of a related photonic work realizing non-Hermitian exceptional topology on a Klein parameter space [64]. While that work focuses on a static scenario and identifies the same-charge EP configuration via spectral Fermi arc measurements, our study reveals and observes an anomalous evolution of EPs enabled by nonorientability, including EP charge inversion, the emergence of HPs and VPs, and a gapped-gapless topological phase transition, with band braids and Berry phases measured as two independent yet complementary invariants.

**Acknowledgements**

This work was supported by the National Key R&D Program of China (Grant No. 2023YFA1406900),

the National Natural Science Foundation of China (Grant Nos. 12374418, 12304495 and 12104346), the Natural Science Foundation of Hubei Province of China (Grant No. 2024AFB654), the Natural Science Foundation of Wuhan (Grant No. 2025040601020223), the Open Project of the Beijing National Laboratory for Condensed Matter Physics (Grant No. 2025BNLCMPKF018), and the Fundamental Research Funds for the Central Universities.